# X-RAY KINEMATOGRAPHY OF TEMPERATURE-JUMP RELAXATION PROBES THE ELASTIC PROPERTIES OF FLUID BILAYERS

Georg Pabst[*], Michael Rappolt[*], Heinz Amenitsch[*], Sigrid Bernstorff[#], and Peter Laggner[*]

[*] Institute of Biophysics and X-ray Structure Research, Austrian Academy of Sciences, Steyrergasse 17, A-8010 Graz, Austria.

[#] Synchrotron Trieste (ELETTRA), SS 14, Km 163.5, I-34012 Basovizza (TS), Italy.

**Corresponding Author:**

Peter Laggner

Institute of Biophysics and X-ray Structure Research, Austrian Academy of Sciences, Steyrergasse 17, A-8010 Graz / Austria

Tel: ** 43 316 812003

Fax: ** 43 316 812367

Email: Peter.Laggner@oeaw.ac.at

# ABSTRACT

The response kinetics of liquid crystalline phosphatidylcholine bilayer stacks to rapid, IR-laser induced temperature jumps has been studied by millisecond time-resolved x-ray diffraction. The system reacts on the fast temperature change by a discrete bilayer compression normal to its surface and a lateral bilayer expansion. Since water cannot diffuse from the excess phase into the interbilayer water region within the 2 ms duration of the laser pulse, the water layer has to follow the bilayer expansion, by an anomalous thinning. Structural analysis of a 20 ms diffraction pattern from the intermediate phase indicates that the bilayer thickness remains within the limits of isothermal equilibrium values. Both, the intermediate structure and its relaxation into the original equilibrium $L_\alpha$-phase, depend on the visco-elastic properties of the bilayer/water system. We present an analysis of the relaxation process by an overdamped one-dimensional oscillation model revealing the concepts of Hooke's law for phospholipid bilayers on a supramolecular basis. The results yield a constant bilayer repulsion and viscosity within Hooke's regime suggesting that the hydrocarbon chains act as a buffer for the supplied thermal energy. The bilayer compression is a function of the initial temperature and the temperature amplitude, but is independent of the chain length.

# INTRODUCTION

Fluid membranes are characterized by only two types of possible elastic deformations: stretching and bending, as the shear modulus within a fluid membrane is zero (for review see Lipowsky, 1995). Experimental methods for resolving the elastic properties of liquid crystalline phospholipid bilayers include a mode analysis of bending fluctuations, pipette aspiration or electric fields induced deformations on unilamellar vesicles (for reviews see, Helfrich, 1995; Evans, 1995). For multilamellar vesicles additional forces arise from interactions between opposing bilayer surfaces. In the case of electrically neutral phospholipids the equilibrium structure is determined by attractive van der Waals forces and repulsive hydration and steric forces (for review see Parsegian and Rand, 1995). The experimental techniques for exploring the interaction potentials are the surface force apparatus, the osmotic stress method, atomic force microscopy and pipette aspiration (Parsegian and Rand, 1995). However, a unifying theoretical description of the liquid crystalline layered structures is still lacking, and presently membrane biophysics relies on a rather empirical exponential model-function of the repulsive forces (Parsegian and Rand, 1995).

Since the elastic properties of phospholipid bilayers determine the equilibrium structure, the bilayer elasticity appears to be even more important for biological non-equilibrium "switching" processes, such as cell fusion or pore formation: only a conservation of the membrane structure will preserve the vital function of the biological membrane in cell compartmentation and communication. Rapid pressure and temperature jump experiments on phospholipid phase transitions (Pressl et al. 1997; Rapp and Goody, 1991) in combination with time resolved x-ray diffraction (Gruner, 1987; Caffrey, 1989; Rapp, 1992; Cunningham et al. 1994) have revealed that the fast phase transitions proceed without any detectable disruption of the multilamellar stack order (Laggner and Kriechbaum 1991; Rapp et al., 1993; Laggner et al., 1999). The most intriguing observation with such experiments is an anomalous, thin lamellar intermediate structure, induced by a T-jump through the pretransition of dipalmitoylphosphatidylcholine (Rappolt M., G. Pabst, G. Rapp, M. Kriechbaum, H. Amenitsch, C. Krenn, S. Bernstorff, and P. Laggner. New evidence for gel-liquid crystalline phase co-existence in the ripple phase of phosphatidylcholines. *Eur. Biophys. J.*, submitted). T-jumps in the $L_\alpha$-phase region have revealed a similar, non-equilibrium intermediate structure – designated as $L_{\alpha*}$ - with a lifetime in the sub-second range (Laggner et al., 1999).

In this article we present an explanation for the observed phenomena of the $L_\alpha$-$L_{\alpha*}$-$L_\alpha$ transition based on a structural calculation of the intermediate structure through a recently developed x-ray analysis method (Pabst, G., M. Rappolt, H. Amenitsch, and P. Laggner. Structural information from multilamellar liposomes at full hydration: X-ray data analysis method. *Phys. Rev. E*, submitted.). We further present a relaxation model which gives qualitative insight into mechanical membrane properties. The results demonstrate that the T-jump technique provides a sensitive way to determine the interacting forces between opposing bilayers and within the bilayer itself, i.e., between the lipid molecules. Thus, the method is an additional tool to gain valuable information on the physical interactions of fluid membrane stacks.

# MATERIALS AND METHODS

## I. Sample preparation

POPC (1-palmitoyl-2-oleoyl-*sn*-glycero-3-phosphocholine), DPPC (1,2-palmitoyl-2-oleoyl-*sn*-glycero-3-phosphocholine) and DSPC (1,2-stearoyl-2-oleoyl-*sn*-glycero-3-phosphocholine) were purchased from Avanti Polar Lipids (Alabaster, AL, > 99% purity) and used without further purification. Multilamellar liposomes were prepared by dispersing weighted amounts of lipids (20 - 30% w/w) in pure water (Fluka, Neu Ulm, Germany: double quartz distilled water, with a specific resistance 18 M$\Omega$ cm) and incubating the dispersions for 4 hours at least 10°C above the main transition temperature to guarantee full hydration. During this period the lipid dispersions were vigorously vortexed. To avoid radiation damage the total exposure-time was kept as low as possible, i.e., each single experiment was carried out with fresh sample but of the same stock. Random thin layer chromatography tests on silica gel-plates 60 (Merck, Darmstadt, Germany) resulted normal. The solvent used was chloroform/methanol/water (65/25/4).

## II. Instrumentation and X-ray Diffraction

Diffraction patterns were recorded at the Austrian Small Angle X-ray Scattering (SAXS) beamline at ELETTRA (Amenitsch et al., 1998; Bernstorff et al., 1998), using a one-dimensional position sensitive detector (Gabriel, 1977) which covered the q-range (q = $4\pi \sin\theta/\lambda$) of interest from about 0.03 Å$^{-1}$ to 0.52 Å$^{-1}$. The angular calibration of the detector was determined by using the SAXS-diffraction patterns of silver-behenate (CH$_3$(CH$_2$)$_{20}$COOAg: d-spacing = 58.38 Å) (Huang et al., 1993). The lipid dispersions were kept in a thin-walled 1 mm diameter Mark capillary held in a steel cuvette, which provides good thermal contact to the Peltier heating unit (Anton Paar, Graz, Austria). T-jumps were generated by an erbium-glass IR-laser (Kriechbaum et al., 1990; Rapp and Goody, 1991). The experimental setup is schematically depicted in Fig. 1. The laser-pulse energy varied from 0.78 to 2.40 J resulting in an average T-jump amplitude $\Delta T$ of 5 to 16°C (Tab. 1). The variation given for the T-jump amplitude derives mainly from the difficulty to measure precisely the energy deposited onto the sample, which depends on the accuracy of measurement of the laser-pulse energy ($\pm$ 5%), the estimation of the geometric properties given in the experimental set-up (10% error), and the estimation of the absorption in the sample (10% error); however, the reproducibility of the laser-pulse energy lies within 2%. A time-resolved experiment (one cycle) consisted of a series of 256 time-frames of pattern detection, with the laser (pulse-length 2 ms), triggered at the beginning of the 11th frame. The maximum time resolution of the X-ray experiment, i.e., right after the laser shot, was set to 5 ms. The total exposure time of one cycle was 16 s. Each experiment was repeated 3 times, i.e., every diffraction pattern was averaged over 3 cycles. The raw data of the time-resolved experiments were normalized for the integration time of each time-frame, and the background (capillary with water) was subtracted.

### III. Structure Calculations

The X-ray diffraction patterns of POPC have been analyzed in terms of the MCG method (Pabst, G., M. Rappolt, H. Amenitsch, and P. Laggner. Structural information from multilamellar liposomes at full hydration: X-ray data analysis method. *Phys. Rev. E*, submitted.). MCG is a diffraction model that combines a modified Caillé theory structure factor (Caillé, 1967; Zhang et al., 1994) with the form factor of a Gaussian representation of the electron density profile such that it accounts for both, Bragg diffraction and diffuse scattering. The method is thus capable of retrieving structural information even if only a few orders of diffraction are observed. The basic concepts of the model are summarized as follows.

The total scattered intensity of stack of lamellae is described by

$$I(q) \propto \frac{1}{q^2}\left(|F(q)|^2 S(q) + N_{diff}|F(q)|^2\right), \quad (1)$$

where $q$ is the absolute value of the scattering vector, $F(q)$ the bilayer form factor given by the Fourier transform of a Gaussian representation of the electron density profile, $S(q)$ the Caillé structure factor and $N_{diff}$ a multiplicator that controls the term for additional diffuse scattering of single, uncorrelated bilayers. Structural parameters are calculated by applying a geometric model, where the bilayer thickness is given by

$$d_B = 2(z_H + FWHM_H/2), \quad (2)$$

with $z_H$ being the position of the headgroup with respect to the bilayer center at the methyl terminus and the FWHM of the Gaussian representing the headgroup in the electron density profile. The interbilayer water layer is the difference of the length of the unit cell minus the bilayer thickness

$$d_W = d - d_B. \quad (3)$$

For further details on the geometric model and MCG we refer to (Pabst, G., M. Rappolt, H. Amenitsch, and P. Laggner. Structural information from multilamellar liposomes at full hydration: X-ray data analysis method. *Phys. Rev. E*, submitted.).

# IV. Relaxation Kinetics

The relaxation kinetics of the lamellar repeat distance d is given by a double-exponential decay (Laggner et al., 1999)

$$d(t) = d_{eq} - d_a \exp(-t/t_a) - d_b \exp(-t/t_b), \qquad (4)$$

where $d_{eq}$ is the equilibrium d-spacing at the given system temperature before the T-jump, $d_a$ and $d_b$ are relaxation components and $t_a$, $t_b$ the respective relaxation time constants. The relaxation process can be compared to a damped oscillation. The differential equation of a one-dimensional damped harmonic oscillator is

$$m\ddot{x} + r\dot{x} + cx = 0, \qquad (5)$$

with the oscillating mass $m$, the friction coefficient $r$, and $c$, the restoring force constant. By solving Eq. 5 one has to distinguish three cases, namely a quasi-periodic motion for slight damping, an aperiodic motion for strong (over-) damping and the aperiodic limit, giving the fastest relaxation to equilibrium without any oscillation. For details on harmonic oscillators see any mathematics texts on differential equations (e.g. G. Joos and E.W. Richter, 1978; C. Schaefer and M. Päsler, 1970). The case of an aperiodic motion gives the solution

$$x(t) = A\exp(-b_1 t) + B\exp(-b_2 t) + C$$

$$b_1 = d + \sqrt{d^2 - w_0^2}$$

$$b_2 = d - \sqrt{d^2 - w_0^2}, \qquad (6)$$

where the frequency $w_0$ is related to the "spring constant" $c$ by

$$w_0 = \sqrt{c/m} \qquad (7)$$

and the damping factor $d$ to the friction coefficient by

$$d = \frac{r}{2m}. \qquad (8)$$

The similar analytic form of Eq. 4 and 6 suggest that the relaxation kinetics of the d-spacing can be seen as a harmonic one dimensional overdamped oscillation. The relaxation time constants $t_a$ and $t_b$ are related to these quantities by

$$w_0 = (t_a t_b)^{-1/2} \tag{9}$$

and

$$d = \frac{t_a + t_b}{2 t_a t_b}. \tag{10}$$

The relaxation components $d_a$, $d_b$ can be expressed by the relaxation velocity

$$v_0 = d_a / t_a + d_b / t_b \tag{11}$$

and the relaxation acceleration

$$a_0 = -d_a / t_a^2 - d_b / t_b^2 \tag{12}$$

at t = 0.

## RESULTS AND DISCUSSION

### I. Equilibrium Structure

We took static X-ray diffraction patterns of POPC liposomal dispersions in a range of 10°C to 70°C within the $L_\alpha$-phase of POPC in excess of water. The sample was held at each temperature for 5 minutes before the measurement was started, so that the system can be regarded as being in thermal equilibrium. Fig. 2 shows the changes in the d-spacing, bilayer thickness, and water layer thickness, as the temperature is increased. The lamellar repeat decreases as the temperature is raised up to 30°C down to a value of d = 63.6 ± 0.1 Å. Above 30°C, the bilayer-water system swells again and finally exhibits a larger lattice parameter at 70°C than at 10°C. The decomposition of the d-spacings into bilayer and interbilayer water thickness reveals that this is caused by an uptake of water, as the membrane thickness continuously decreases with increasing temperature, but the bilayer separates more and more, such that the sum of both gives the observed re-increase in d-spacing. With respect to the results which will be presented in subsection III, we draw the attention to the membrane thickness, which first decreases linearly with temperature, but shows an asymptotic behavior above 50°C. The characteristics of the $L_\alpha$-phase are "molten" hydrocarbon chains due to a transition from a all-*trans* to a *trans-gauche* state. An increase in temperature progressively induces *trans*-to-*gauche* conformation changes – the minimal energy difference between *trans* and *gauche* states is 500 cal mol$^{-1}$ (Flory, 1969) –, but since the hydrocarbon tails are finite in length, only a limited number of *gauche* isomers can be generated. This corresponds to a saturation effect in hydrocarbon chain melting as has been found before, e.g., for DPPC (Seelig and Seelig, 1974) that causes the asymptotic behavior of the membrane thickness, which is also observed for POPC above 50°C (Fig. 2). The reduction of the bilayer thickness has been first described by Luzzati and co-workers, and explained in terms of a rubber-like model for the hydrocarbon chain elasticity (Luzzati, 1968). Concomitantly with the decrease in thickness, the bilayer responds to the temperature increase by a lateral expansion, as each single phospholipid molecule requires more space due to the increasing degrees of motional freedom at higher temperatures (for review see Cevc and Marsh 1987; Hianik et al., 1998). The linear thermal expansion coefficient ***a***, defined as

$$\boldsymbol{a} = \frac{\Delta d_B}{d_B} \frac{1}{\Delta T} \tag{13}$$

is given as a function of temperature in Fig. 3. The mean value of $\alpha = -2.2 \cdot 10^{-3}$ K$^{-1}$ below 50°C is close to the linear thermal expansion coefficient of $-2.7 \cdot 10^{-3}$ K$^{-1}$ for K-soaps (Luzzati, 1968) and $-2.5 \cdot 10^{-3}$ K$^{-1}$ for DPPC (Seelig and Seelig, 1974). The thermal expansion coefficient decreases linearly from 50°C to 70°C to a value of $a = -0.5 \cdot 10^{-3}$ K$^{-1}$, indicating a reduced bilayer elasticity. At first sight, this appears to be surprising as one would expect a more fluid bilayer at higher temperatures with increased disorder. However, fluid DPPC bilayers have been found to have less order, but a ten times larger viscosity than soaps, e.g., sodium decanoate-decanol bilayers (Seelig and Seelig, 1974).

## II. Intermediate Structure

Fig. 4 shows the kinematographic diffraction patterns of a typical jump/relaxation experiment. The structure of the $L_\alpha$-phase under equilibrium conditions at the given system temperature $T_0$ is the starting point and reference for the T-jump experiment (Fig. 4). With the laser flash (duration: 2 ms) the original structure converts quasi discontinuously, i.e., within the time resolution of the experiment[*], into a lamellar phase, whose d-spacing is clearly thinner than the corresponding lattice parameter under equilibrium conditions at $T = T_0 + \Delta T$ (Laggner et al., 1999). The excited phase, denoted as $L_{\alpha*}$, relaxes within the time scale of seconds back to the equilibrium d-spacing. The FWHM of the $L_{\alpha*}$-phase is smaller than the corresponding values of the original $L_\alpha$-phase.

In the following, the intermediate structure is to be compared with the equilibrium structure obtained in the previous subsection. As a reference we choose the structure of POPC bilayers at 20°C, with a d-spacing of 63.9 ± 0.1 Å, a bilayer thickness of 46.7 ± 0.3 Å, and a bilayer separation of 17.2 ± 0.3 Å (Fig. 2). The laser voltage was set to 600 V, which corresponds to a temperature jump of 10 ± 2°C (Tab. 1). In this particular experiment, the T-jump/relaxation cycles were repeated 12 times, and the single diffraction patterns were added up to reduce statistic noise. We further summed up the first four 5 ms-diffraction patterns after the laser pulse, during which the d-spacing relaxes by approximately 0.3 Å only, again to improve the statistics of the $L_{\alpha*}$ diffraction pattern. The structural analysis on $L_{\alpha*}$ has been performed by applying the MCG model (Pabst, G., M. Rappolt, H. Amenitsch, and P. Laggner. Structural information from multilamellar liposomes at full hydration: X-ray data analysis method. *Phys. Rev. E*, submitted.), yielding 62.1 ± 0.5 Å for the lamellar repeat distance, i.e., by 1.5 Å thinner than the minimum repeat distance found under equilibrium conditions (Fig. 2), 45 ± 1.7 Å for the bilayer thickness, and a water layer of 17.1 ± 1.7 Å thickness. The corresponding bilayer thickness at 30°C under equilibrium conditions is 45.6 ± 0.3 Å, and the value of $d_w$ is 18 ± 0.5 Å (Fig. 2). Within the limits of measurement error it can be assumed that the bilayer shrinks proportionally to the heat deposited, according to

$$d_{B*} = d_B(T_0 + \Delta T), \tag{14}$$

wherein $d_{B*}$ denotes the bilayer thickness of $L_{\alpha*}$. We can further calculate the change in interbilayer water volume by

$$\Delta V_W = A^* d_W^* - A d_W, \tag{15}$$

where $A$ and $A^*$ are the respective areas per phospholipid molecule, which can be estimated according to the formalism of Nagle and co-workers (Nagle et al., 1996). For the given temperature of $T_0 = 20°C$, we calculate a volume change of $\mathbf{D}V_W = -17$ Å$^3$. For comparison: the volume of one water molecule is approx. 30 Å$^3$. Thus, the change in interbilayer water volume can be regarded as zero. However, we estimate the error to be larger than 50 Å$^3$, since the employed estimates of the T-jump amplitude and of the area per lipid include uncertainties, which strongly influence the result.

---

[*] The diffraction pattern between the well defined $L_\alpha$ and $L_{\alpha*}$-phase is recorded during the heating of the laser pulse. The broad diffuse peak of this pattern is due to the fast changes in the lattice parameter and not to a less ordered phase.

The result predicts a anomalous thin water layer of the intermediate structure: the temperature-induced decrease in membrane thickness is accompanied by a lateral expansion of the bilayer (Cevc and Marsh, 1987). As there is no water exchange with the excess aqueous phase directly after the laser shot ($DV_W = 0$), the interbilayer water will follow the lateral bilayer expansion which gives the observed thin layer. The transient water deficit can also explain the observed sharp Bragg peaks of the $L_{\alpha*}$-phase, as more layers can contribute coherently to Bragg diffraction, similarly to phosphatidylethanolamine multilayers, which incorporate less than half the amount of water compared to phosphatidylcholine multilayers, and which exhibit sharper Bragg peaks than phosphatidylcholines in the $L_\alpha$-phase (McIntosh and Simon, 1986).

### III. Relaxation Kinetics

The relaxation kinetics of the IR-laser T-jump induced $L_\alpha$-$L_{\alpha*}$ transitions have been analyzed in terms of the relaxation model presented in the theory section (Eqs. 4, 9-12). Fig. 5a shows the changes in the lattice parameter as a function of time for a 16°C T-jump from $T_0 = 20$°C. Time equals zero at the first frame after the laser shot. The individual data points were obtained by fitting the first order Bragg peak of each single diffraction pattern by a Lorentzian function. The solid line in Fig. 5a gives the best fit of the d-spacing relaxation to the double exponential decay model (Eq. 6) yielding a fast time of constant $0.45 \pm 0.02$ s and a slower component of $t = 3.0 \pm 0.1$ s. The lower part of Fig. 5a further illustrates the time course of the maximum intensity and FWHM of the first order Bragg peak. The intermediate phase, at t = 0, has a thinner FWHM and an increased peak intensity as compared to the original $L_\alpha$-phase. Both, maximum intensity and FWHM, decay rapidly to values below the equilibrium $L_\alpha$ reference values and start to recover after 2 s. Even after 16 s, when the d-spacing has retained its equilibrium value, the peak intensity and FWHM indicate that the relaxation process is still going on. The total relaxation time is about 30 to 40 s. Fig. 5b depicts the first and the second derivative of the d-spacing relaxation, i.e., the relaxation velocity and the relaxation acceleration, for the first 5 seconds of the relaxation process. The relaxation is initially driven by a strong acceleration which diminishes after 2-3 seconds, whereupon the relaxation proceeds with constant velocity.

We shall try to obtain further insight from energetic aspects. Fig. 6 depicts the interaction potential between two bilayer surfaces. For neutral phospholipids the interacting forces are attractive van der Waals, repulsive hydration, and steric forces (Marra and Isrealachvili, 1985; Lipowsky, 1995; Parsegian and Rand, 1995). The bilayer separation is given by the minimum of the potential curve. At higher temperatures, the potential well is shallower, and the minimum moves to larger bilayer separations. Through the T-jump, the bilayer/water system is quasi discontinuously forced into the intermediate state (Fig. 6 ⊗), sensing the strong repulsive part of the interaction potential that drives the water layer thickness towards the equilibrium separation (Fig. 6 ⊕). As the temperature of the systems does not remain at $T_0 + \Delta T$, but slowly decreases to $T_0$, the interaction potential changes continuously, and the equilibrium separation moves again to smaller values. The bilayer separation has to follow the changes via a series of potential wells.

So far, we have considered the bilayer/water system as a closed system that does not interact with the excess water phase. In reality, also diffusion of bulk water through the phospholipid bilayers will have an impact on the relaxation process. Diffusion proceeds either directly through the membrane or through local membrane defects, i.e., several times faster, and is driven by the hydration forces due to the water deficit. Thus, the rate of water transport into the interbilayer space will affect the relaxation velocity by retarding (damping) the relaxation process.

Since the temperature of the heated sample volume decreases in the later stages of the experiment by heat diffusion into the unheated sample regions, we have to consider two processes, at the same time: a relaxation – due to the anomalously thin water layer – and a bilayer swelling – due to the decrease in temperature. Both processes are governed by the elastic properties of the bilayer/water system. However, we expect the first process, i.e., the structural relaxation, to be initially dominant over the temperature dependent bilayer swelling. The rapid drop in intensity (Fig. 5a) and the relaxation acceleration (Fig. 5b) within the first 2-3 seconds after the laser pulse can be therefore interpreted as the fingerprint of this relaxation process. The constant relaxation velocity (Fig. 5b) is then attributed to the second – swelling – process. The swelling velocity is governed by the interplay of temperature dependent molecular rearrangements within the phospholipid bilayer and water diffusion in and out of the water layers, as the system will generally face both non-equilibrium positions on the interaction potential well (Fig. 6), attracting and repelling. A possible relaxation pathway of the bilayer separation is sketched in Fig. 6.

A further interesting aspect of T-jump experiments in the $L_\alpha$-phase is the relaxation behavior of the lattice parameter as a function of starting temperature ($T_0$), and temperature jump amplitude ($\Delta T$), respectively. We first present the results upon varying the starting temperature. The liposomal dispersion of POPC was equilibrated at temperatures between 10 and 70°C. At each temperature we performed a T-jump experiment with a jump amplitude of $\Delta T = 10 \pm 2$°C. The relaxation kinetics of the d-spacings were analyzed in terms of the double exponential decay model (Eqs. 4, 9 – 12). Fig. 7 shows the results for the most important parameters, i.e., the change in repeat distance $\mathbf{D}d$, the square of oscillation frequency $\mathbf{w}_0^2$, the damping factor $\mathbf{d}$, the zero velocity $v_0$ and the zero acceleration $a_0$. As $T_0$ is increased, the absolute value of $\mathbf{D}d$ decreases linear up to a temperature of 50°C. This agrees well with the findings under equilibrium conditions, where the membrane thickness also exhibited a linear decrease in the range of 10°C to 50°C (Fig. 2). For the present purposes, water can be regarded as an incompressible medium, and therefore, $\mathbf{D}d$ is directly related to the compressibility of the phospholipid membrane stacks.

The results therefore demonstrate that below 50°C, the transmitted thermal energy can be directed into a membrane thinning. Mechanically, this regime corresponds to a linear relationship between applied tension and deformation, i.e., Hooke's law. This is also expressed in the constancy of the other parameters $\mathbf{w}_0^2$, $\mathbf{d}$, $v_0$ and $a_0$. However, these parameters cannot be attributed to membrane properties only. Here, the bilayer/water system has to be considered as an entity, since it is the interplay of repulsive and attractive forces of bilayer/water interactions which governs the relaxation process and thus influences these parameters. As described in the theory section, the square of the oscillation frequency $\mathbf{w}_0^2$ is directly related to the proportionality constant of the repelling forces, which are dominated by hydration potentials (Fig. 6). The results depicted in Fig. 8 clearly show that, as long as the linearity of Hooke's law is given, the relaxation proceeds with the same "spring constant". This further emphasizes that the transmitted energy can be transformed into hydrocarbon *trans-gauche* transitions within the regime of Hooke's law, so that the bilayer separation and hence also the bilayer repulsion after the laser pulse is practically the same. In this sense, the hydrocarbon chains act as a buffer for the thermal energy.

A similar behavior is found for the damping factor, which is related to the system inertia or viscosity. As the viscosity of water can be neglected, the *d* parameter is a measure of viscosity of the phospholipid bilayers. The viscosity also includes the permeability of the membrane for water, since we concluded before, that water will diffuse from the excess phase into the water layer of the membrane stacks. Above 50°C, the total change in d-spacing exhibits saturation, no further compression of the bilayer/water system can be achieved with the supplied laser energy. As found under equilibrium conditions, the bilayer is less elastic above 50°C due to the maximum number of *gauche* isomers reached. The transmitted energy cannot be completely buffered by the bilayer, so that the reaction, i.e., membrane thinning, to the sudden temperature increase is not strictly linear, resulting in a minimal constant *Dd* above 50°C. The excess of thermal energy results in a stronger repulsion of the membrane surfaces as observed in the increase of $w_0^2$, *d*, $v_0$ and $a_0$.

Fig. 8, depicts the changes in the relaxation parameters as a function of the temperature jump amplitude. The sample was equilibrated at 40°C and the laser voltage was varied from 500 to 750 V in steps of 50 V, yielding the different T-jump amplitudes (Tab. 1). With increasing T-jump amplitude, the absolute value of *Dd* increases linear. The square of the oscillation frequency and the damping factor show a statistic variation, but can be regarded as constant within in the measurement error. On the other hand, both zero velocity and zero acceleration show a strong statistic variation, but a tendency to increase with increasing T-jump amplitude. At 40°C, the system is very close to the end of validity of Hooke's law. This proximity to non-linearity may be the reason for the strong statistic fluctuations of the relaxation parameters $w_0^2$, *d*, $v_0$ and $a_0$. Nevertheless, the results clearly state that the discrete jump in the repeat distance *Dd* is a linear function of the transmitted laser energy and that the other

relaxation parameters are independent of the T-jump amplitude, as long as the linearity of Hooke's law holds.

If T-jumps are sensible to the hydrocarbon chain elasticity, then one would expect different results for phospholipids with different chain lengths. Table 2 shows the fitted relaxation parameters for fluid DPPC, POPC and DSPC multilayers. The laser energy was adjusted to a T-jump amplitude of 16°C. The results reveal that the discrete change in d-spacing $\Delta d$ is within the measurement error for all three lipids equal to ~ 2.6 Å. Since this shrinkage can be attributed to hydrocarbon chain melting, we conclude that an equal amount of energy deposit causes an equal amount of *trans-gauche* transformations. However, the other parameters exhibit a dependence on the lipid type. Most interestingly, the square of the oscillation frequency is highest for POPC, such that the increasing order with respect to $w_0^2$ is DPPC < DSPC < POPC. This might be an effect of the unsaturated 18:1c9 chain of POPC, but as $w_0^2$ is related to the repelling forces, the reason might also be a better packing of the phospholipids at lower temperatures (Tab. 2), because a more compact bilayer will provide a better surface for the affecting repulsive forces. Nevertheless, the damping coefficient $\delta$ increases with hydrocarbon chain length, which is reasonable as one would expect a higher viscosity for thicker membranes. The zero velocity $v_0$ and zero acceleration $a_0$, respectively, exhibit the same dependence on hydrocarbon chain length.

# CONCLUSION

The reaction of the liquid-crystalline multilayer system to the T-jump proceeds via an excited state. This intermediate structure $L_{\alpha*}$ is characterized by a membrane thickness which compares to its equilibrium value at $T_0 + \Delta T$ and an anomalously thin water layer, as the result of a lateral expansion of the membrane and the impossibility of a sufficiently fast water diffusion from the excess phase into the water layers of the membrane stacks. Strong repulsive hydration forces then initiate the relaxation process (Fig. 5), where the membrane viscosity and its permeability for water controls the relaxation velocity. As the heat within the sample volume dissipates into the surrounding unheated sample, the relaxation proceeds over a series of potential wells and the finial state is equal to the original $L_\alpha$-phase (Fig. 6). Fig. 9 shows a cartoon of the T-jump induced structural transformation; the POPC bilayer has been generated by a molecular dynamics simulation (Heller et al., 1993). Since the only two types of elastic deformations for fluid membranes are stretching and bending (Lipowsky, 1995), the possible forms of the deformed intermediate liposomes are elongated vesicles or liposomes with a rippled "harmonica-like" topology (left inset of Fig. 9).

In addition to a structural model for the $L_\alpha$-$L_{\alpha*}$ transition, the T-jump/relaxation experiments were found to provide a qualitative, fundamental insight into the elastic properties of the phospholipid bilayers. The elastic properties of model membranes have been discussed ever since Luzzati and co-workers described the hydrocarbon chain properties as rubber-like (Luzzati, 1968), where the impact of the elasto-mechanic membrane features is of high interest for membrane protein interactions (e.g. Mouritsen and Bloom, 1984; Nielsen et al., 1998, Lundbæk and Andersen, 1999). The standard measurement techniques for mechanical bilayer properties are pipette aspiration, the surface force apparatus, the osmotic stress method and atomic force microscopy (Evans, 1995, Parsegian and Rand, 1995). However, the generic interactions of stacks of fluctuating membranes are still missing an accepted theoretical formulation, i.e., the physical basis of the hydration forces is not yet understood. Here T-jump experiments can give additional information, since they directly probe the membrane stacks mechanic characteristics as has been demonstrated in the present work for three different phospholipids (Tab. 2). So far the concept relies on a qualitative analysis given by the relaxation parameters $Dd$, $w_0$ and $d$. Future research will be directed towards a relaxation theory that allows for a quantitative evaluation of membrane properties. As physical interaction prediction is of prime importance also for liposomal based rational drug design or nano-materials, the T-jump/relaxation method shall open the door to a wider field of applications.

TABLE 1 Summary of the laser-pulse energy chosen and the resulting temperature amplitudes ΔT.

| Applied Laser Voltage (V) | Energy of the laser-pulse (J) | Temperature-jump amplitude ΔT (°C) |
|---|---|---|
| 500 | 0.79 ± 0.04 | 5 ± 1 |
| 550 | 1.15 ± 0.06 | 8 ± 1 |
| 600 | 1.42 ± 0.07 | 10 ± 2 |
| 650 | 1.76 ± 0.09 | 12 ± 2 |
| 700 | 2.04 ± 0.10 | 14 ± 2 |
| 750 | 2.41 ± 0.12 | 16 ± 2 |

TABLE 2 Relaxation parameters for three phosphatidylcholines (all 20% w/w) at a T-jump amplitude of 16°C.

|  | DPPC | POPC | DSPC |
|---|---|---|---|
| chains | 16:0/16:0 | 16:0/18:1c9 | 18:0/18:0 |
| $T_0$ (°C) | 50 | 30 | 70 |
| $\Delta d$ (Å) | 2.5 ± 0.3 | 2.8 ± 0.3 | 2.5 ± 0.3 |
| $\omega_0^2$ (s$^{-2}$) | 0.044 ± 0.002 | 1.1 ± 0.1 | 0.34 ± 0.02 |
| $\delta$ (s$^{-1}$) | 0.72 ± 0.03 | 1.6 ± 0.1 | 2.6 ± 0.2 |
| $v_0$ (Å/s) | 1.5 ± 0.1 | 5.0 ± 0.7 | 7.7 ± 0.7 |
| $a_0$ (Å/s$^2$) | - 2 ± 0.4 | - 13 ± 2 | - 40 ± 2 |

# Figure Captions

FIGURE 1 Experimental set-up of a T-jump experiment at the SAXS-beamline (Elettra). The IR-beam of a Erbium glass laser is directed via a prism onto the sample capillary which is thermostated by a Peltier heating unit. The high flux x-ray beam transverses the sample normal to the IR-beam. The structural changes are recorded by a position sensitive detector with a maximum time resolution of 5 ms.

FIGURE 2 The equilibrium structure of POPC bilayers in the $L_\alpha$-phase at different temperatures. The changes on d-spacing, membrane thickness and interbilayer water thickness are depicted. The observed re-increase in lamellar repeat distance is due to a uptake of water which increases the bilayer separation, whereas the membrane itself gets thinner with temperature and exhibits an asymptotic behavior above 50°C.

FIGURE 3 The elastic properties of POPC bilayers at equilibrium. The linear expansion coefficient $\alpha$ exhibits a constant value up to 40°C, and reduces linear as the temperature is increased further.

FIGURE 4 The $L_\alpha$-$L_{\alpha*}$-$L_\alpha$ transition in a liposomal dispersion of POPC induced by a 16°C temperature jump. The series of time-sliced diffraction shows the time course of the first order Bragg reflections. The laser is triggered at t = 0.

FIGURE 5 The relaxation kinetics of the $L_\alpha$-$L_{\alpha*}$-$L_\alpha$ transition. (a) The temporal evolution of the lamellar repeat distances, obtained by a Lorentzian fit to the first order Bragg peaks. The straight line gives the best fit of the relaxation model Eq. 4 to the data points. The lower part

depicts the changes in the Bragg peak intensity (▲) and FWHM (○) during a T-jump relaxation experiment. (b) The relaxation velocity and the relaxation acceleration during the first 5 seconds after the laser shot.

FIGURE 6 Illustration of the interaction potential as a function of bilayer separation at different temperatures. The IR-laser shoots the system into a situation (⊗) far away from equilibrium, where it faces the repulsion of hydration forces. As the temperature does not remain at $T_0 + \Delta T$, but decreases with time to $T_0$ the equilibrium position (⊕) at the reached temperature will in general not be reached.

FIGURE 7 The relaxation parameters of POPC liposomes as a function of initial temperature $T_0$. The T-jump amplitude was 16°C at each temperature.

FIGURE 8 The relaxation parameters of POPC liposomes as a function of the T-jump amplitude. The samples were equilibrated at $T_0 = 40$°C before each T-jump experiment.

FIGURE 9 Cartoon of the T-jump induced structural chances on fluid bilayers (using PDB-files by Heller et al., 1993). An increase in temperature induces *trans-gauche* transitions in the fatty acid tails and an increase in lateral area per each single phospholipid molecule (bottom right insert). This leads in the compound of the membrane to a compression normal to the bilayer surface and a lateral bilayer expansion. Since water cannot diffuse fast enough from the excess phase into the interbilayer water region the water layer thickness reduces to an anomalous thin value. The bottom left insert depicts the possible intermediate forms of the liposomes.

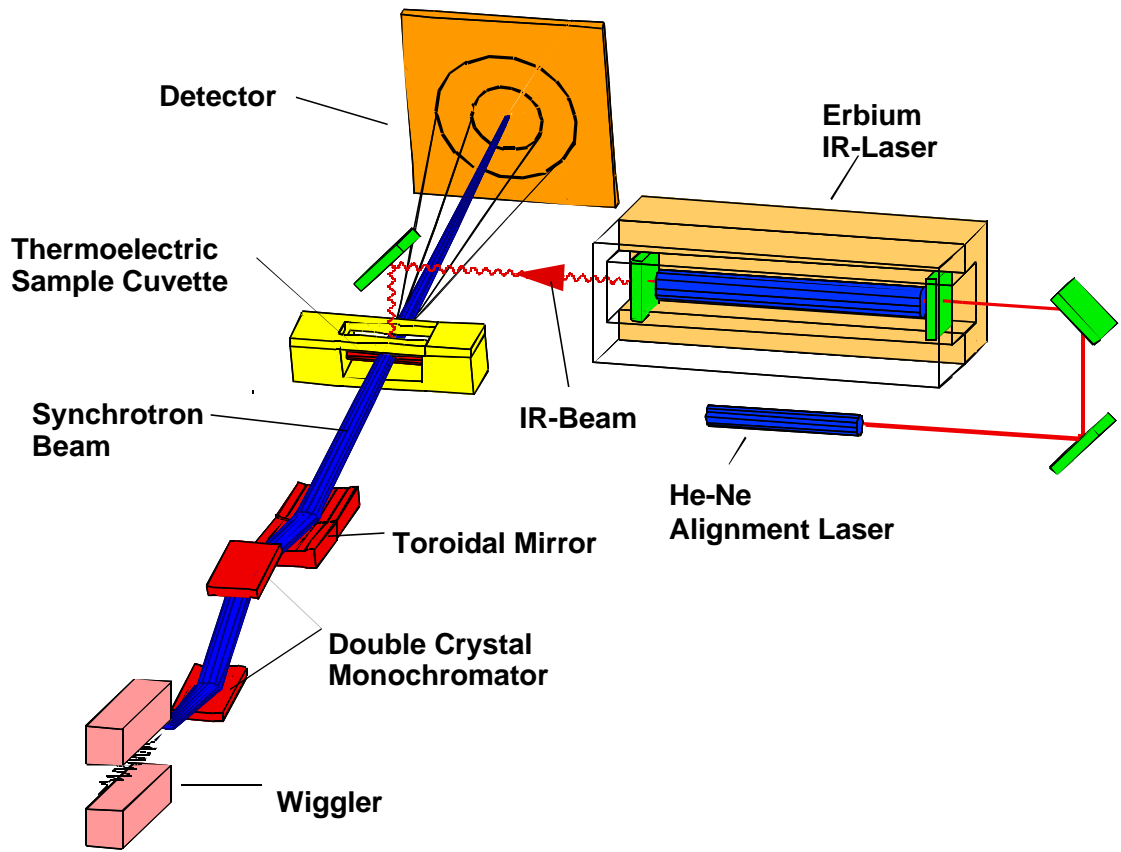

Pabst / Fig. 1

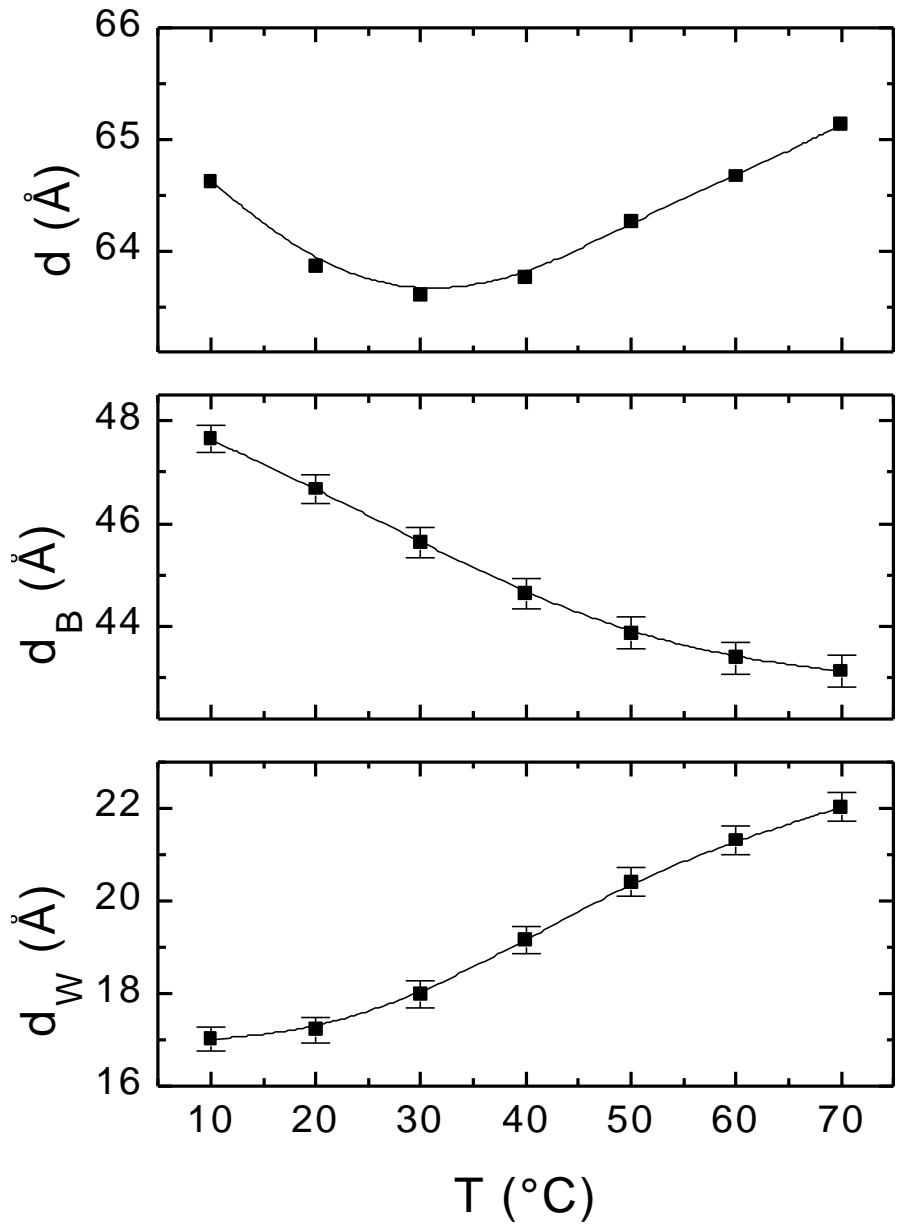

Pabst / Fig. 2

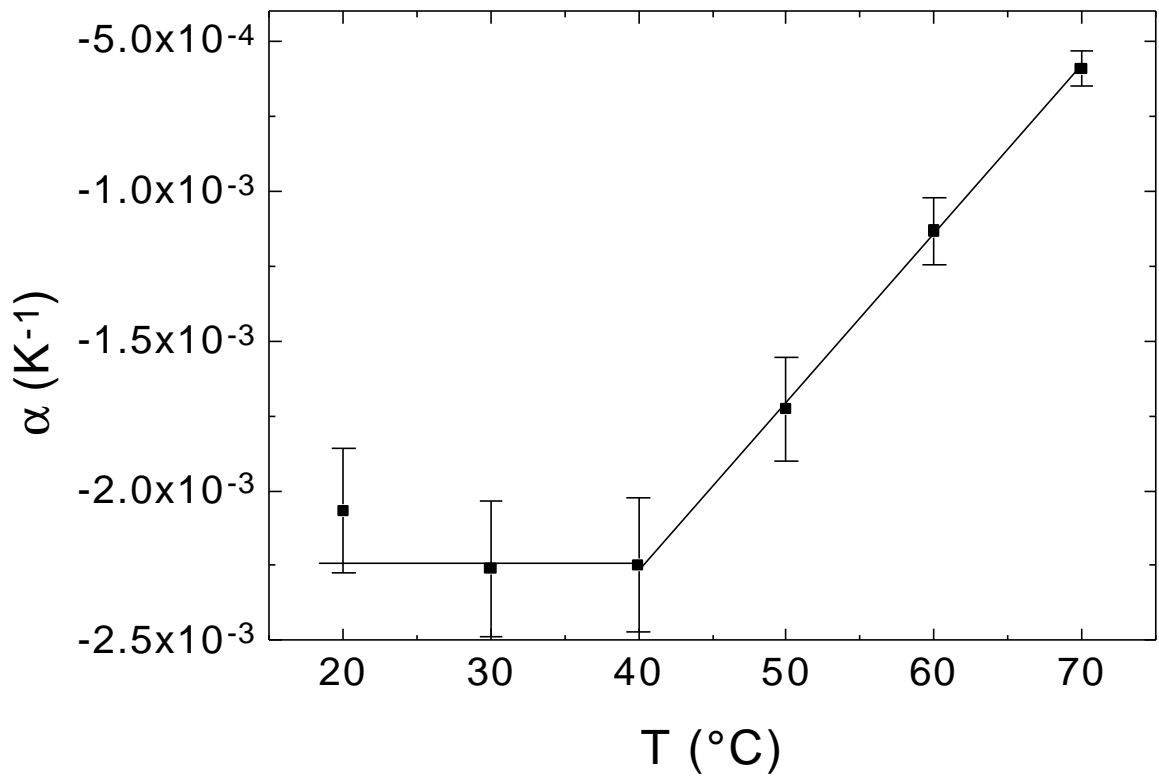

Pabst / Fig. 3

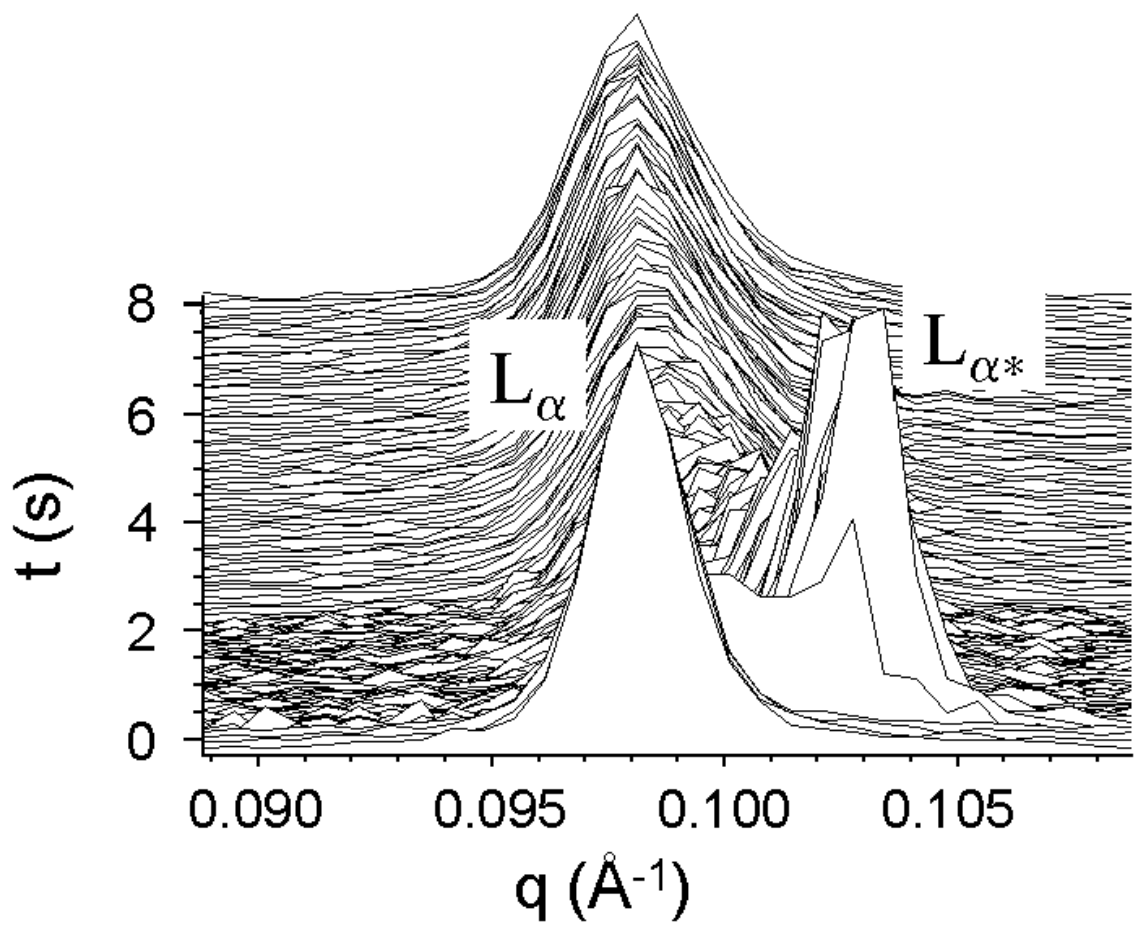

Pabst / Fig. 4

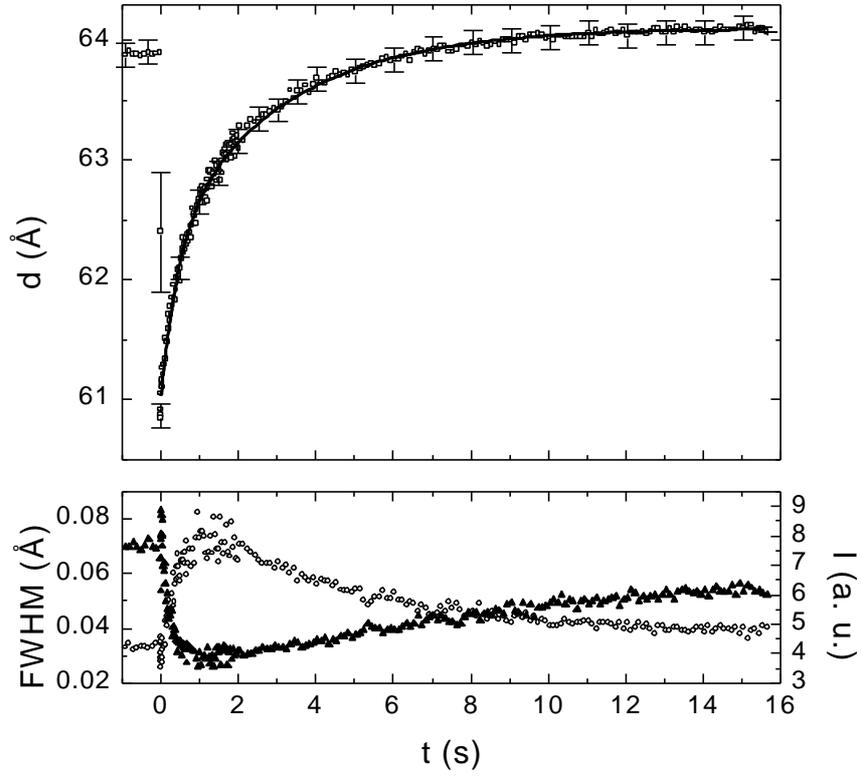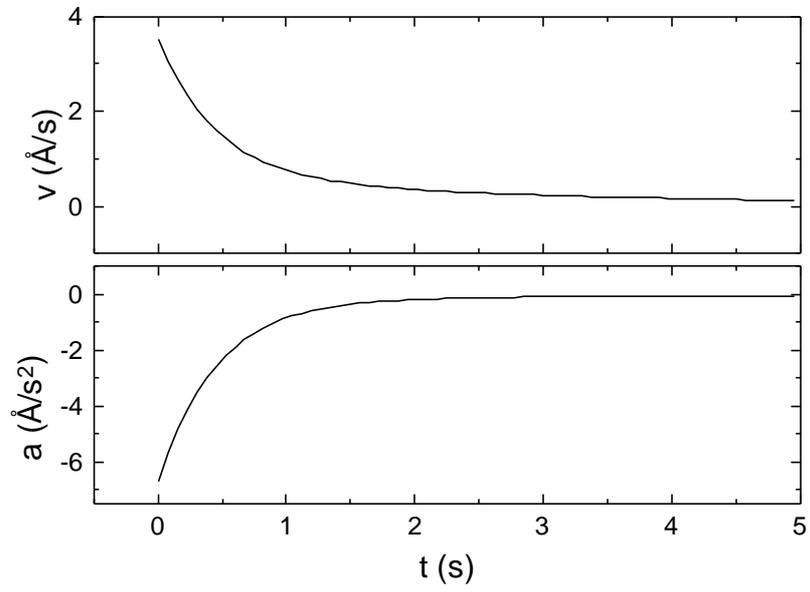

Pabst / Fig. 5

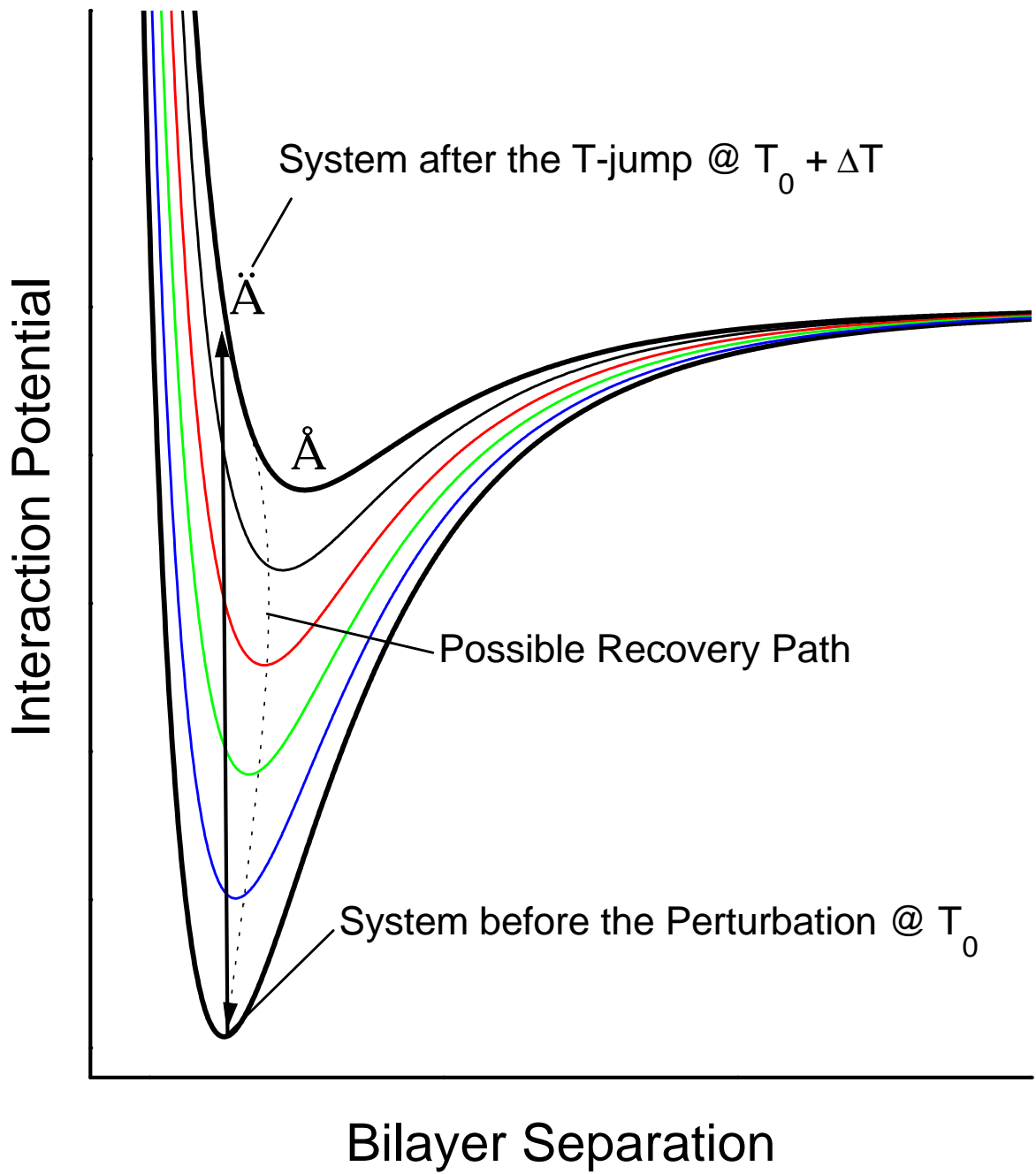

Pabst / Fig. 6

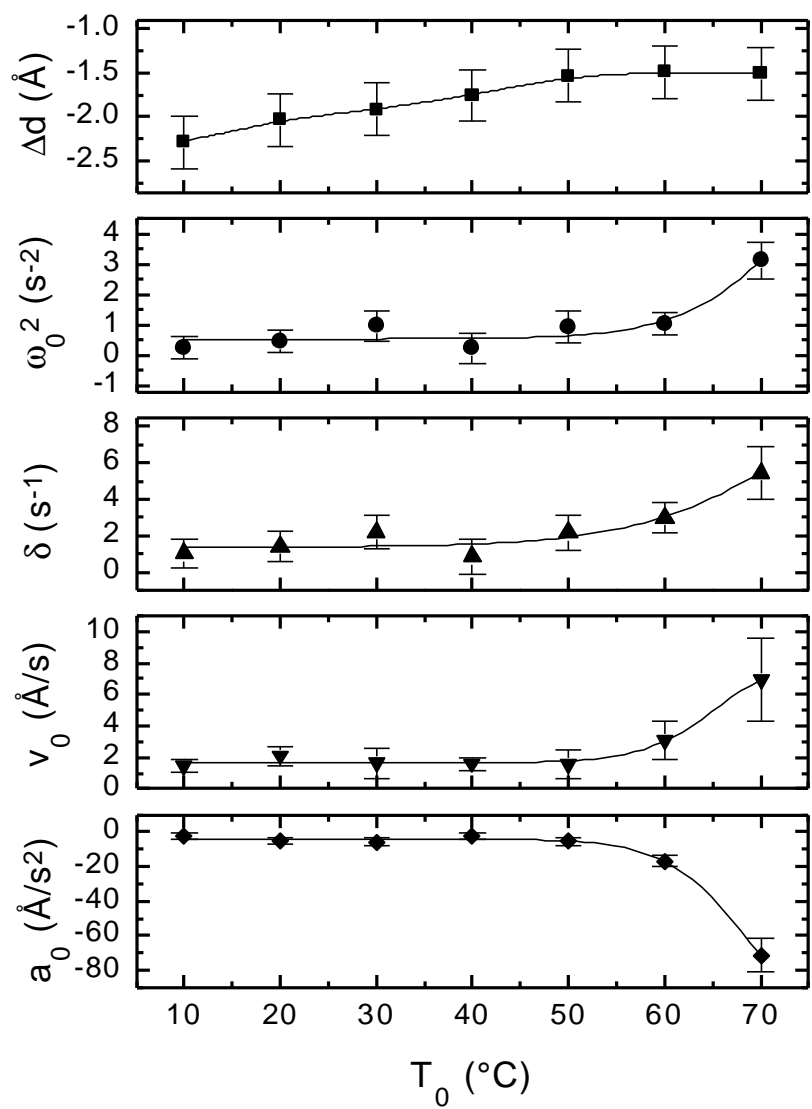

Pabst / Fig. 7

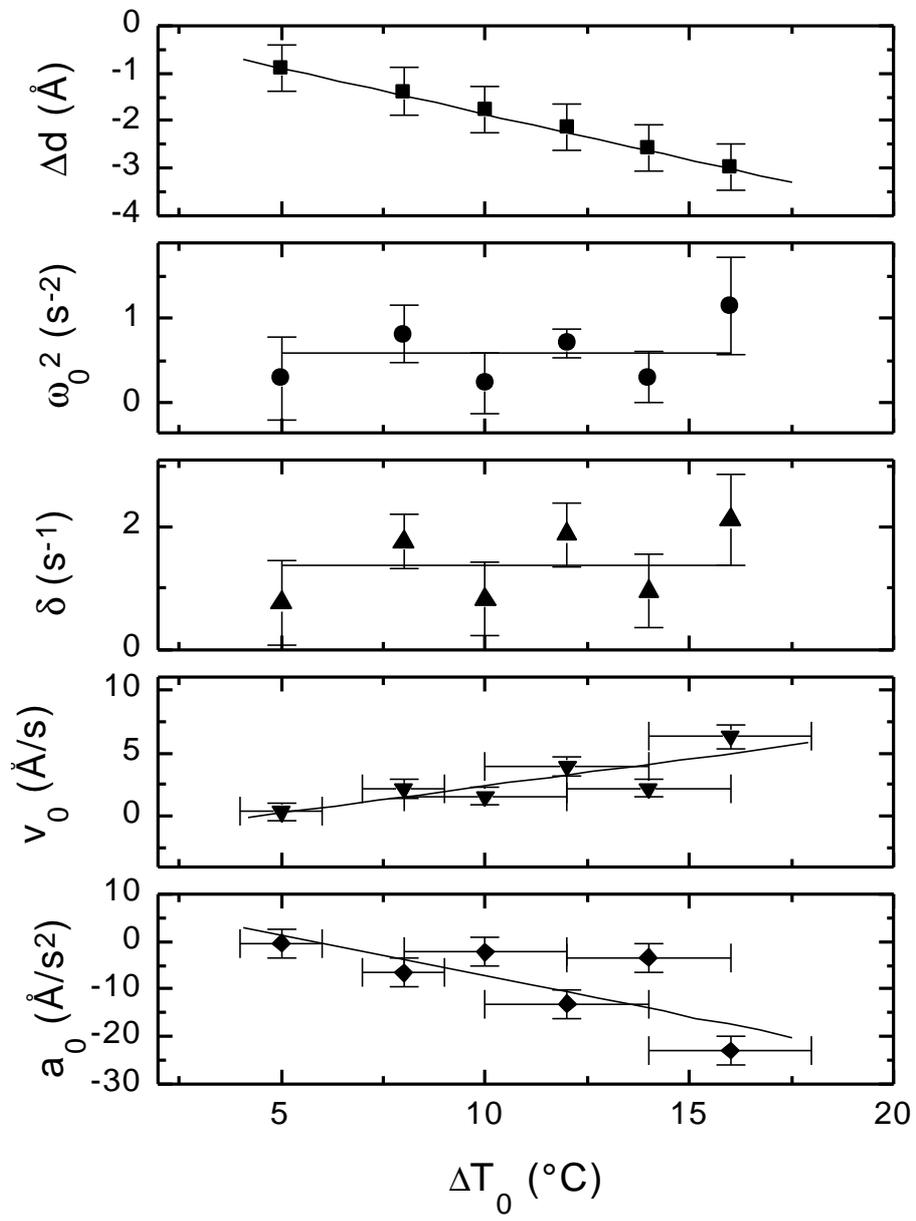

Pabst / Fig. 8

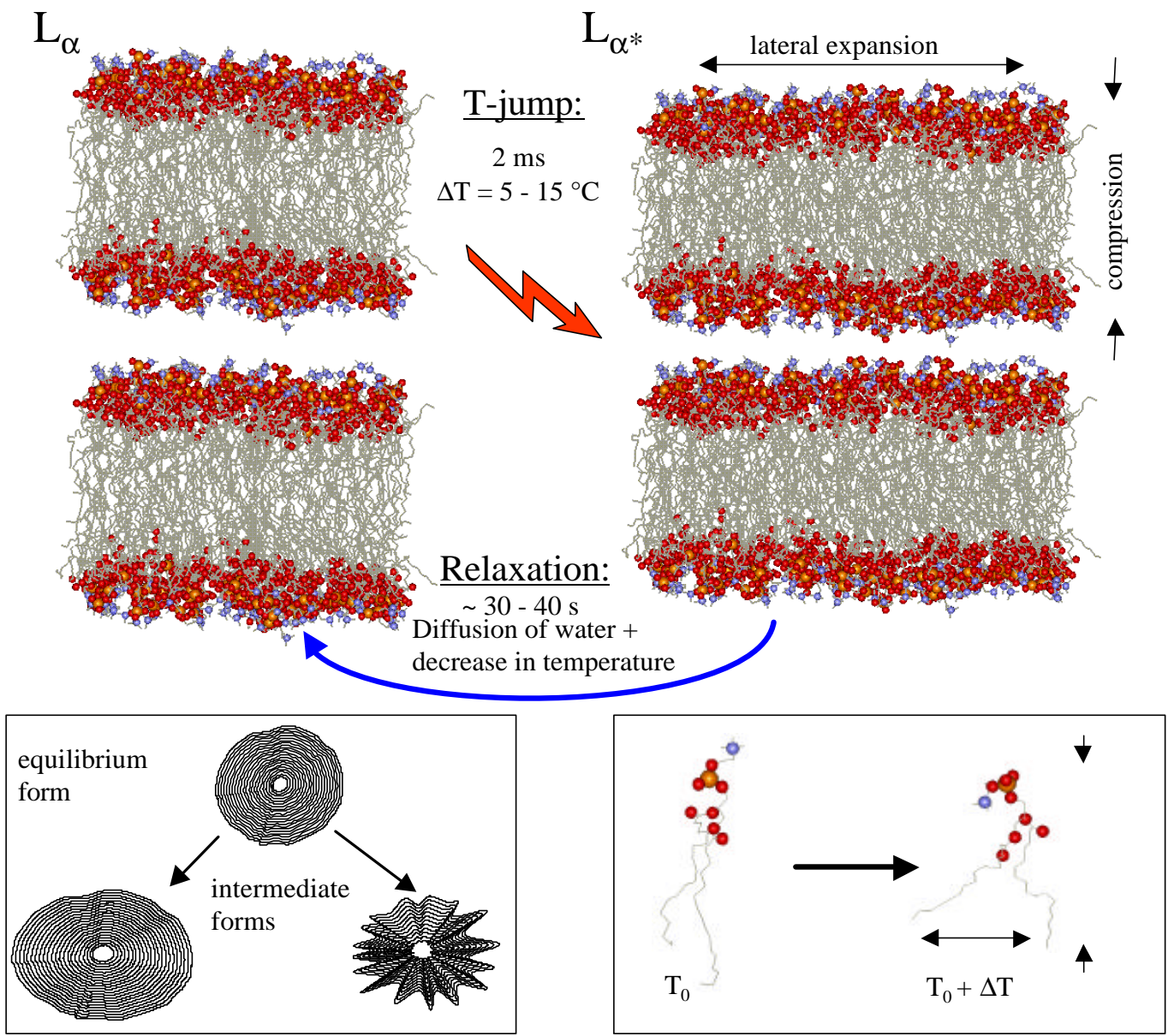

Pabst / Fig. 9